\documentclass[a4paper,10pt]{article}
\usepackage{txfonts}
\usepackage{graphicx}	
\usepackage{url}	
\title{%
Developing Cloud Chambers with High School Students}

\author{Ryo Ishizuka, Nobuaki Tan, Shoma Sato and Syoji
Zeze$^*$\\
Yokote Seiryo Gakuin High School, Akita 013-0041, Japan \\ \\
$*$ ztaro21@gmail.com
}

\begin{document}
\maketitle
\begin{abstract}
 The result and outcome of the \textit{cloud chamber project},
which aims to develop a cloud chamber useful for science education
is reported in detail.  A project includes both three high school
students and a teacher as a part of Super Science High School (SSH)
program in our school.  We develop a dry-ice-free cloud chamber 
using salt and ice (or snow).  Technical details of the chamber are
described.  We also argue how the project have affected student's
cognition, motivation, academic skills and behavior.
 The research project has taken steps of professional researchers,
i.e., in planning research, applying fund, writing a paper and 
giving a talk in conferences.   From interviews with students, 
we have learnt that such style of scientific activity is very effective
in promoting student's motivation for learning science.
\end{abstract}

\section{Introduction}

Since the Fukushima nuclear accident on March 2011, the requirement
of teaching radiation has been significantly increasing in Japan.  
However, recent surveys\cite{ref,hosaka1,hosaka2,fukui} show that
experiments or measurements concerned with radiation are rarely done 
in a classroom.  In order to improve such a situation, an easy, 
safe and inexpensive experiment will play a key role.
One may agree that the diffusion cloud chamber experiment\cite{
langsdorf,needels,cowan} meets
most requirements for the purpose mentioned above. 
Indeed, it is reported\cite{ref} that
the cloud chamber experiment is most popular among radiation related
experiments.  

While the experiment is simple enough,  there may also be some inconveniences
which prevent a teacher from bringing it into a classroom.  One such 
inconvenience is the {\it use of  dry ice}.  Even in Japan, there are
many cities where obtaining dry ice is not so easy.
Even if available, dry ice can not be kept for a long time.  
A teacher has to order dry ice day by day for a long term use.   
To overcome this difficulty, various \textit{dry-ice-free} experiments 
have been developed \cite{Kamata, ice}.  
In this article, we would like to explore the possibility of
using ice and salt to cool a cloud chamber.  The remarkable advantage 
of our method is that both ice and salt are easily available and can be 
stored for a long time. Moreover, they are very familiar to us.  
Our city, Yokote is famous for traditional snow dome called {\it Kamakura},
therefore is abundant in snow.  A particular kind of salt is 
spread on the winter roads as road salt to avoid freezing.   
In fact, authors used to wonder whether snow can be used 
for cloud chamber while doing experiments using dry ice or liquid
nitrogen. This strong motivation leads us to develop a cloud chamber
using ice and salt.
\begin{figure}[htbp]
\begin{center}
\includegraphics[bb=0 0 192 288,scale=0.5]{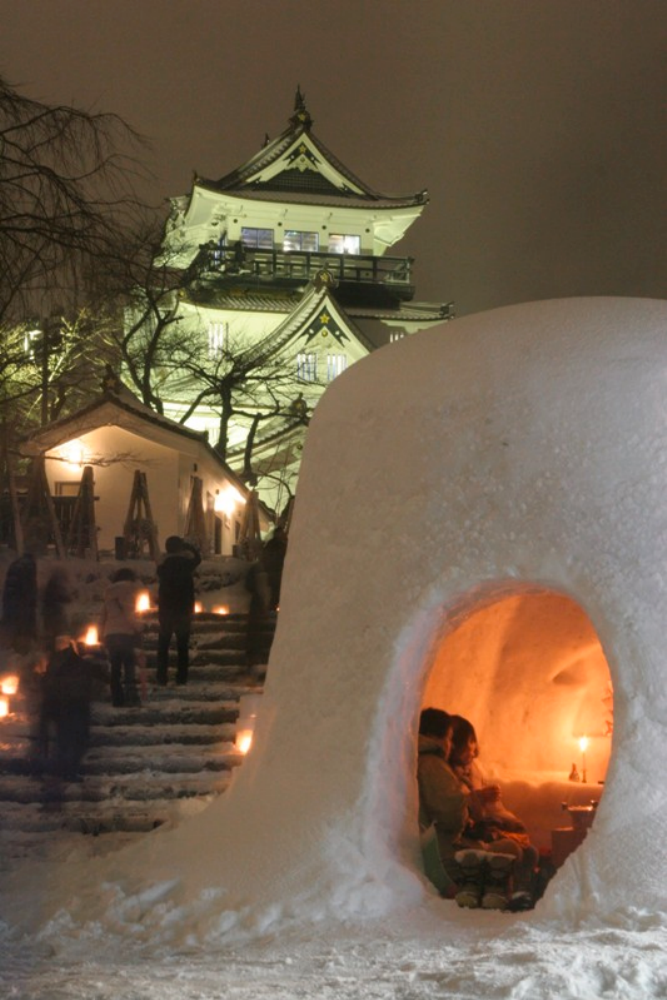} 
\caption{A snow dome \textit{Kamakura}.}
\label{012554_5Jul13}
\end{center}
 \end{figure}

This paper also aims to identify educational effects of
student research project.  Our project lunched January 2011 as 
an activity of science club, with support from the Super Science
High School (SSH) program in our school.  The team consists of a teacher (S.Z.) 
and three high school students who are also authors of this paper.  
Students first determined their research subject, 
applied for some research funds,  wrote a paper and 
presented their result in more than five conferences. 
A presentation in an international conference in Taiwan was given in English.
Providing a realistic environment for research can be regarded as an
example of situated learning proposed by Lave and Wenger \cite{LPP}.  
Such strategy is applied in most SSH schools in Japan, 
since its effectiveness to motivate students and to develop scientific
skills has become evident.  A survey \cite{Ogura} reported that student's motivation 
for learning science is significantly enhanced by an experience
of a spontaneous research project organized by themselves.  It was also
reported that an experience of research affects students
careers.\cite{Ito}  We would like to identify changes in students, ideas about
science, career and scientific research through interviews with
students.   
\begin{table}[htbp]
\caption{List of conference talks}
\label{062800_4Oct13}
\begin{tabular}{lll}
\hline
 Name & Place  & date
\\ \hline
 SEES2012 & I-shou university, Taiwan  & Aug.~2012\\
 Science conference & Akita university & Oct.~2012\\
 Tohoku SSH conference & Sendai 3rd High School & Feb.~2013\\
 JSPS junior session & Hiroshima university & Mar.~2013 \\
 SSH conference & Yokohama & Aug.~2013  \\
\hline
\end{tabular}

\end{table}

\section{A cloud chamber using ice and salt}

\subsection{Ice-salt cooling bath}

\begin{figure}[htbp]
\begin{center}
\includegraphics[bb=0 0 432.96 180.48, scale=0.5]{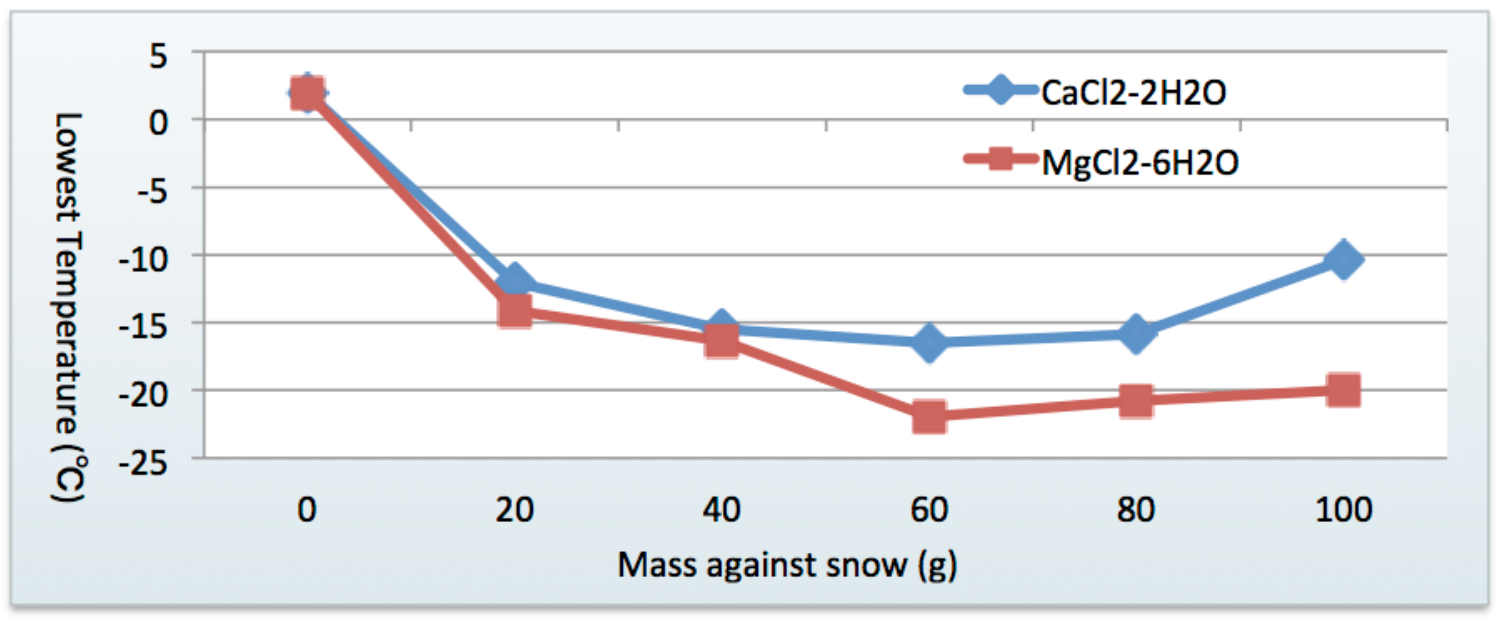} 
\caption{Performance evaluation of two salts.}
\label{saltcurve}
\end{center}
 \end{figure}

It is well known that  table salt melts ice and the mixture of
salt and ice (the cooling bath) becomes colder after  salt added. 
This phenomena, freezing point depression, is common to various
salts other than table salt (sodium chloride).  Our first mission
is to find a  better salt which shows greater performance in cooling 
than a table salt. We choose two salts, \textit{calcium chloride
dihydrate} ($\mathrm{CaCl_{2}.2H_{2}O}$) and 
magnesium chloride hexahydrate ($\mathrm{MgCl_{2}.6H_{2}O}$) as our
candidates, since they are known to perform 
better than table salt \cite{coolingmix} 
and are also available at low cost as road salts. Figure \ref{saltcurve}
compares the performance of two salts for cooling.  The experiment is done
by adding a particular amount of salt to 100g of crushed ice in 
a polystyrene foam tray.  Measurements were done for ten minutes for each
amount of salt to record lowest temperature reached. 
It turns out that $\mathrm{MgCl_{2}.6H_{2}O}$ shows better performance 
for cooling than $\mathrm{CaCl_{2}.2H_{2}O}$. 
The best performance is obtained when we add 60g of 
$\mathrm{MgCl_{2}.6H_{2}O}$, in which the temperature of the mixture
reached $-22\ {}^\circ\mathrm{C}$.

\subsection{Construction of a chamber}

After some trials for constructing  a cloud chamber for 
an ice-salt cooling bath,
we learned that the design of the bottom plate for cooling
is not straightforward.   First, heat capacity of
bottom plate should be small since the ice-salt slush
is not as cold as dry ice.  It is very difficult to see particle tracks
unless the chamber is neatly designed.   Second, the 
bottom plate and the slush should be kept in contact.
If one simply put a chamber onto
slush, it floats on the slush and becomes unstable.  
\begin{figure}[htbp]
\begin{center}
\includegraphics[bb=0 0 500 246,scale=0.5]{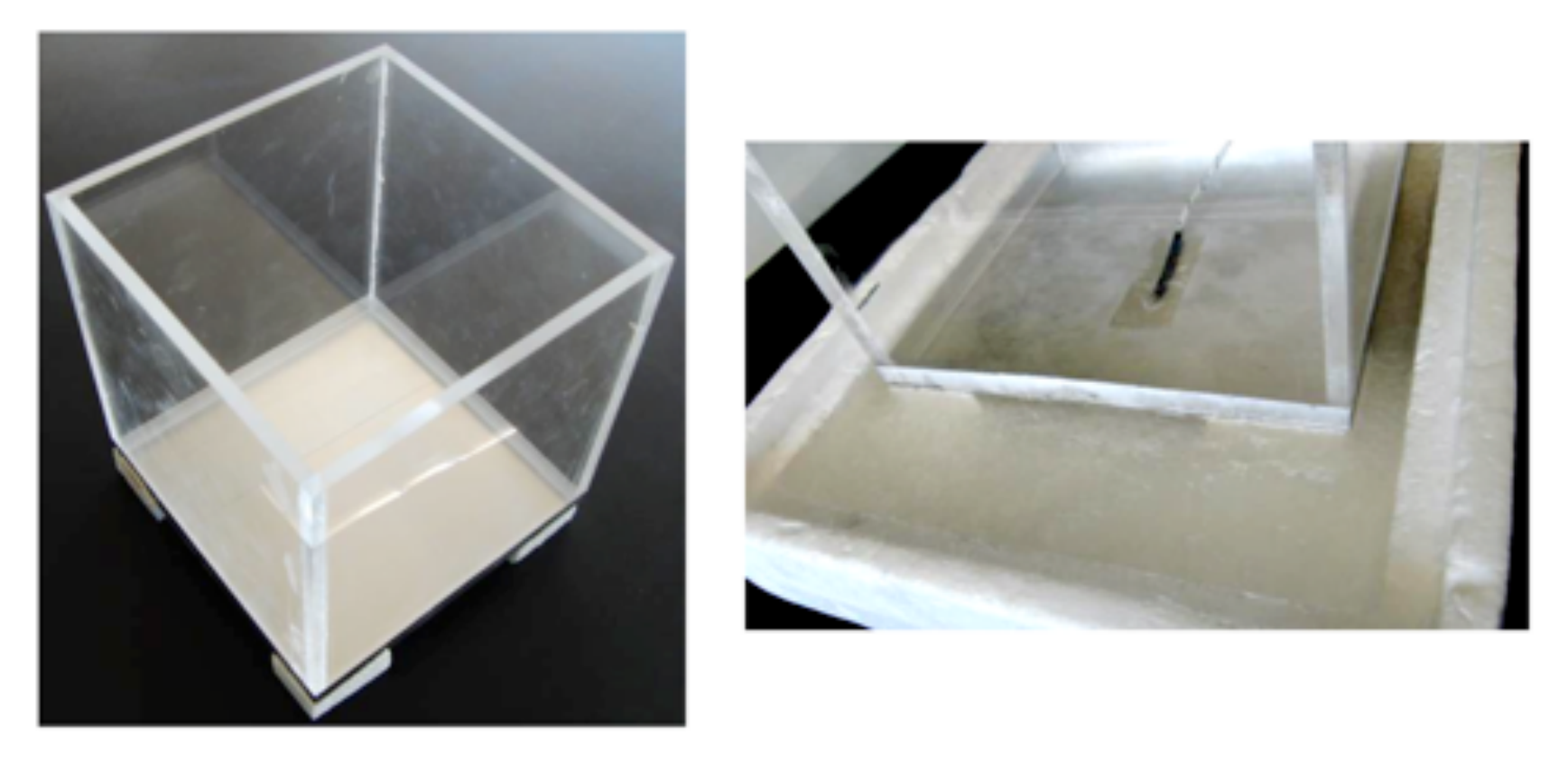} 
\caption{The bottom of the cloud chamber.}
\label{bottom}
\end{center}
 \end{figure}
To overcome these problems, we designed a cloud chamber as
Fig. \ref{bottom}. The bottom plate is made of thin aluminum.  In order
to avoid the chamber float, legs are attached at four corners of the
bottom plate.  As shown in right side of Fig. \ref{bottom}, we pour 
the ice-salt slush around chamber so that it is kept in contact with
the bottom plate.   

\subsection{Observation of the tracks}

We would like to describe an observation of particle tracks
using the ice-salt cooling bath ($\mathrm{MgCl_{2}.6H_{2}O}$ is employed) 
and the chamber described above. 
 \begin{figure}[htbp]
\begin{center}
\includegraphics[bb=0 0 410 374,scale=0.4]{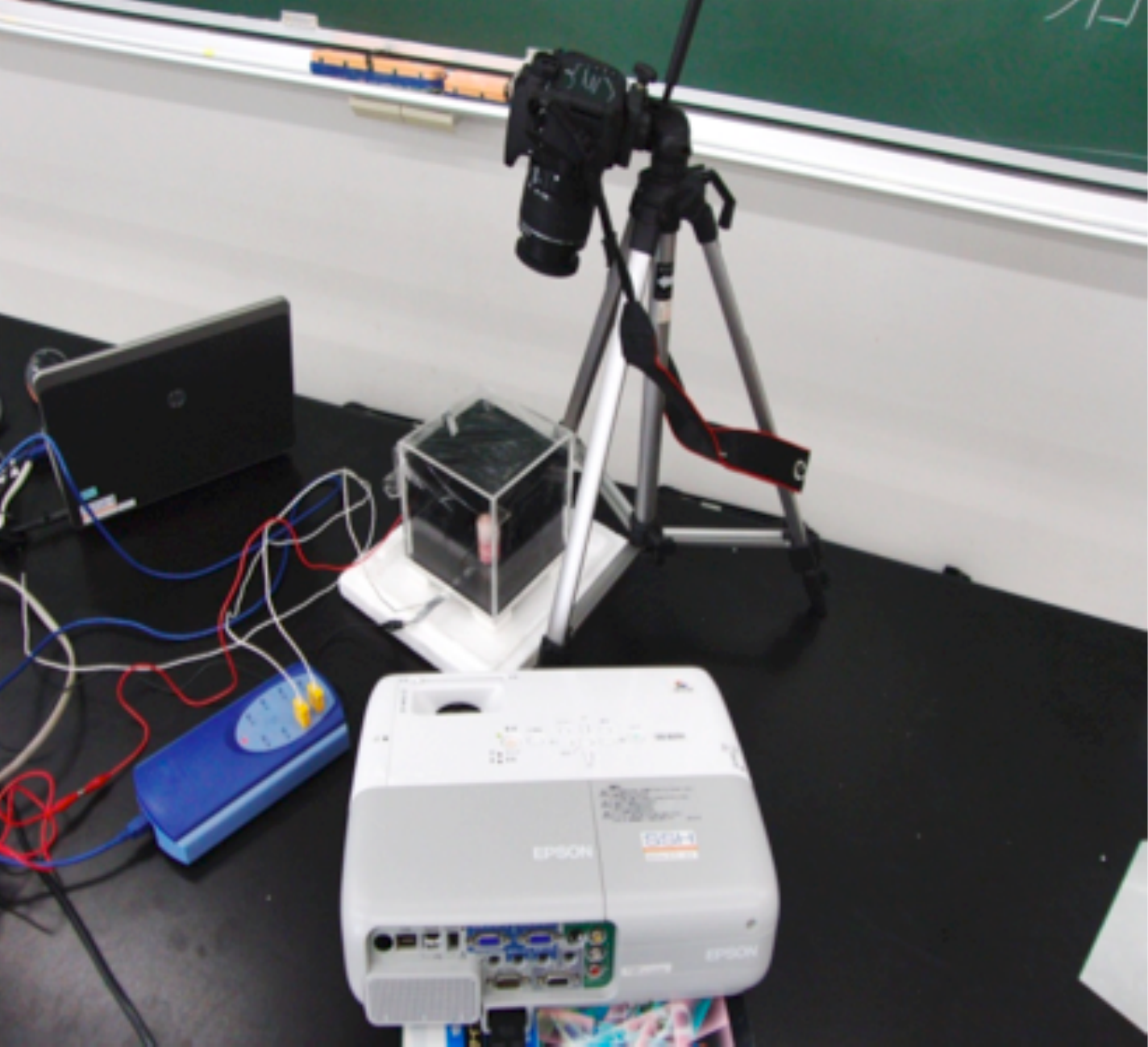} 
\caption{The whole apparatus.}
\label{whole}
\end{center}
 \end{figure}
The whole apparatus is shown in Fig. \ref{whole}.  
The interior of the chamber is covered by a piece of 
black felt soaked with ethanol.  
A video projector is used as a strong light source.  
Temperature of some places inside the chamber is measured by a
thermocouple.  A lantern mantle is placed in the chamber
as an alpha particle source.  A digital SLR camera is used to
record video of particle tracks. The room temperature was around 
$25{}^\circ\mathrm{C}$.  

\begin{figure}[htbp]
\begin{center}
\includegraphics[bb=0 0 271.68 185.28, scale=0.6]{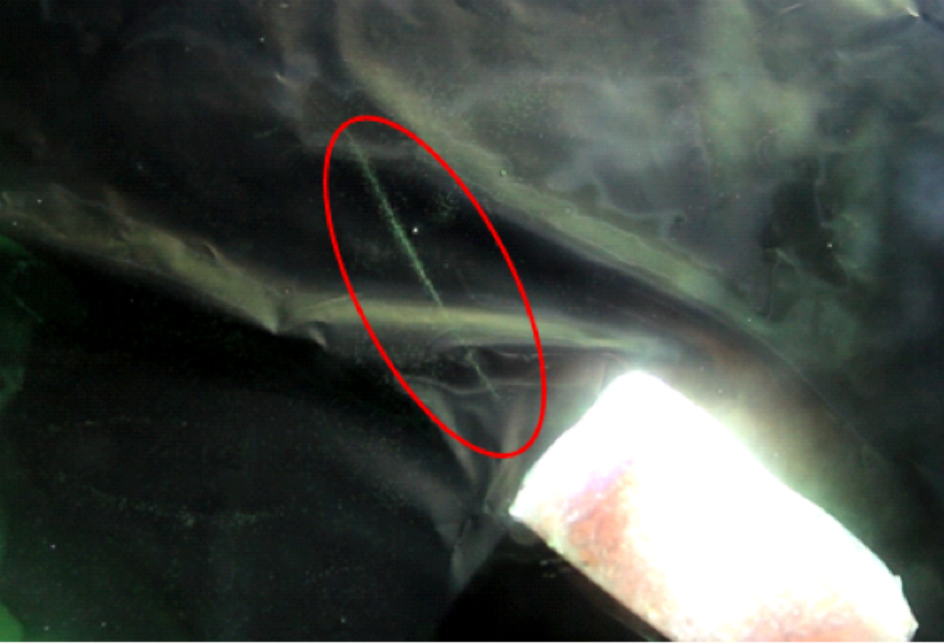} 
\caption{An alpha particle track.}
\label{track}
\end{center}
 \end{figure}

\begin{figure}[htbp]
\begin{center}
\includegraphics[bb=0 0 417.6 271.68, scale=0.6]{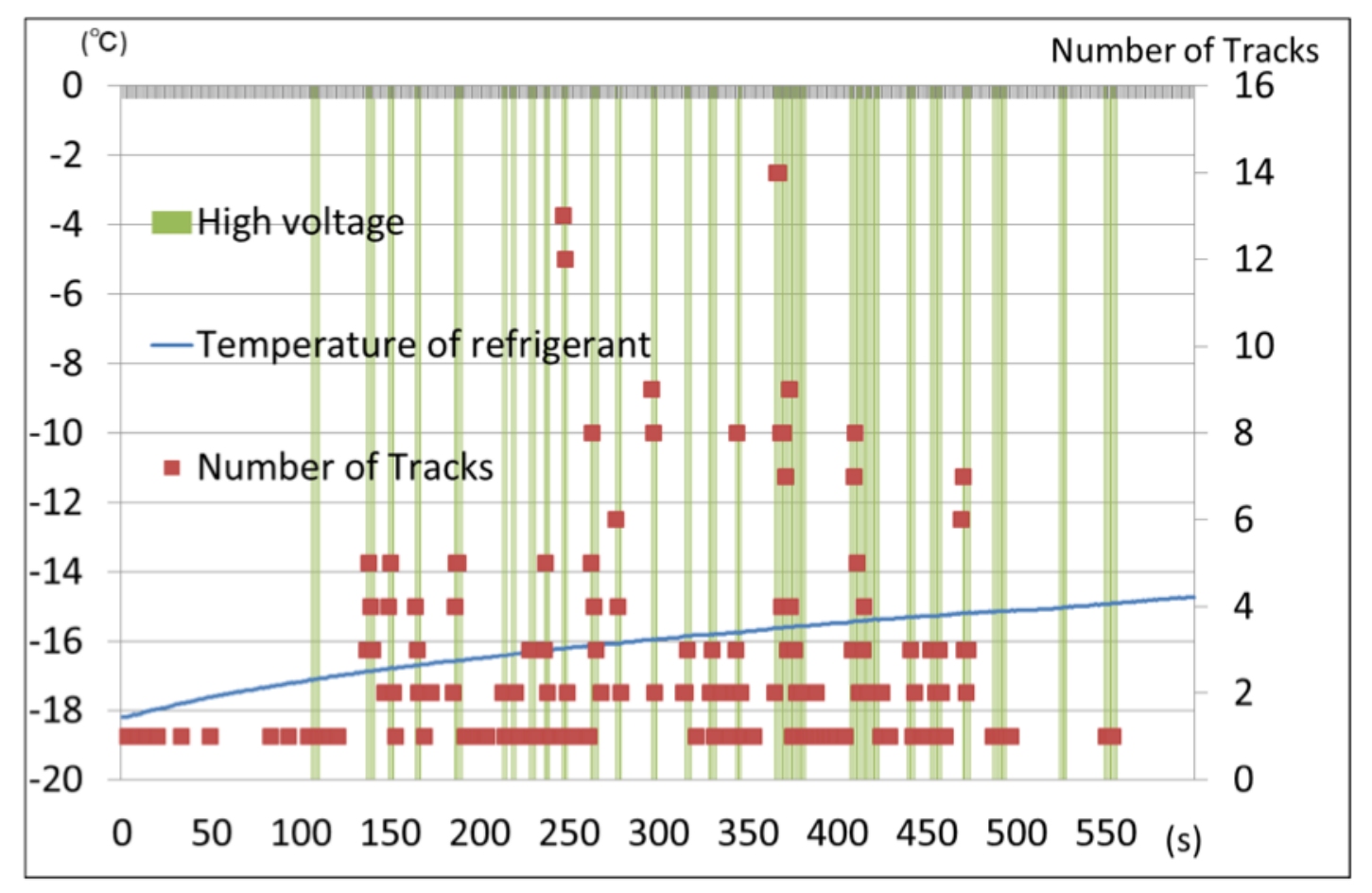} 
\caption{The count of particle tracks.  }
\label{count}
\end{center}
 \end{figure}

With this setting, we succeed in observing particle tracks as seen in
Fig.~\ref{track}, although the number of tracks 
is quite less than that of a dry ice experiment.
A result of more quantitative analysis using a video image is shown in
Fig. \ref{count}, in which a 10 minutes run is recorded.  A track longer
than 2cm is counted.  41 tracks are observed per minute.  
The vertical line in figure shows a period in which 
we apply voltage by approaching charged plastic ruler from the outside
of the chamber.  Apparently a count increases when voltage is applied. 
It is also remarkable that tracks remain visible even when the bottom
of the chamber rises up to $-16{}^\circ\mathrm{C}$.

\section{Educational effects on students}

\subsection{History of the research project}

As mentioned in the introduction, our project begins as a part of 
SSH program of our school which is largely 
supported by MEXT and JST.  An inquiry based class (100 minutes per
week, 1 year) and science club activity (an hour per day, 1.3 years ) were spent for research.
In accordance with the very high requirement of SSH,
students had worked hard to accomplish their research goal.  In
our case, the goal is to observe particle tracks with the ice and salt 
chamber.  It was very close to the deadline of Japan Student Science
Award (JSSA), which is one of the most important scientific award in Japan, 
that they succeeded in observing particle tracks.  

The teacher (S.Z.) assumed that, it is very important not to give
detailed instruction to students in order to make their activity 
spontaneous.  Actually,  most parts of the project were conducted by
themselves.  The teacher gave minimum instructions for experimental
technique (use of liquid nitrogen) and theoretical issue (vapor
pressure).  In particular, the theory of droplet nucleation is far beyond 
the high school physics curriculum.  It is surprising  that they
learned it themselves and plotted the vapor pressure after minimum
instruction of Mathematica.  
\begin{figure}[htbp]
\begin{center}
\includegraphics[bb=0 0 720 960, scale=0.2]{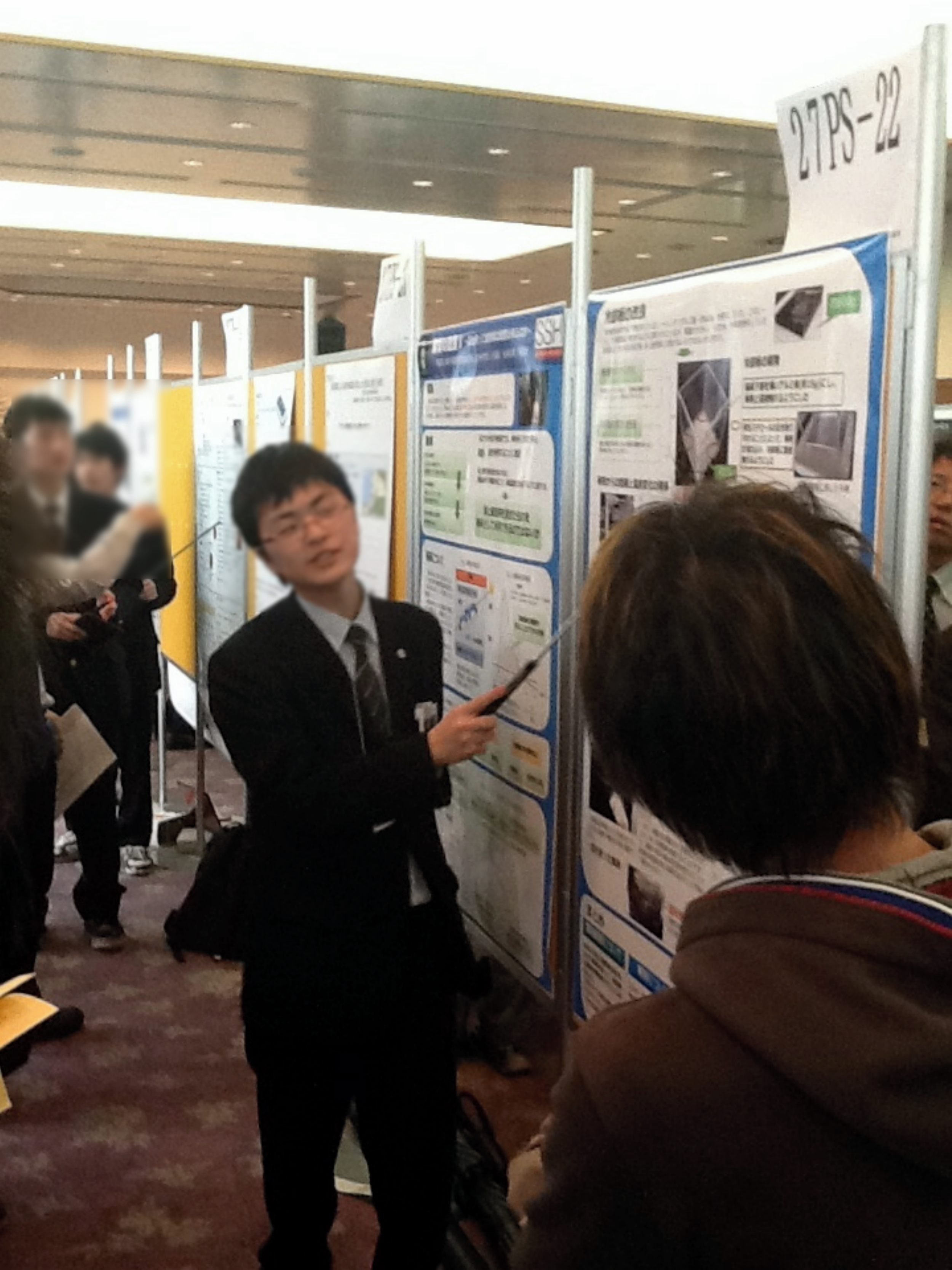} 
\caption{A poster talk  in JPS junior session  on Mar. 2013}
\label{JPS}
\end{center}
 \end{figure}
Among conference talks listed in Table \ref{062800_4Oct13},  
the poster presentation in junior session 
of Japanese Physics society was very impressive. 
Their effort to give the best presentation was rewarded by both a 
good presentation award and
also responses from researchers.   We did not expect many
responses from physicists since the cloud chamber is no longer used as a 
real research tool for particle physics.  It is surprising that our
presentation drew interests not only from physics education researcher but
also experimentalists.  A professor kindly informed us of a B.S. thesis
that is not published.   Through this experience, we have learned the
importance of having opportunity for communicating with other researchers
to obtain up-to-date information and forming a community for particular
research topic.  Such opportunities largely depend on the supoort from
SSH  for travel expenses, which are not available in ordinary school.

\subsection{Interviews}

The strategy of the research project, providing environment and 
opportunities as close as possible to a graduate student physics 
lab, is common for SSH Schools in Japan.  
According to the literature \cite{stipek, Ogura},  
motivation for learning and mastering
science increases if what students do is under their control.   
In this section, we would like to present a result of an interview
with three students in order to evaluate educational effects of
our project.  Interviews 
was carried out from 9 to 12 July, 2013.  Each students spent about 20 minutes 
to answer seven questions.  The questions are:
\\
\\
\noindent
\fbox{
  \parbox{.45\textwidth}{
\small 
\begin{enumerate}
 \item Can you find any differences between usual lectures in 
      science class and our project research?
 \item Have you grown up through this project? If so, in what sense?
 \item Do you think that you did your own project which is not 
assigned or forced from our school?
 \item Have you accomplished something great? 
 \item Did you find any difficulties to accomplish your mission? 
 \item Did you find any difference between peers or teachers in your school
and professional researchers met in conferences ? 
 \item What do you think about speaking, reading and writing English in
       our project? 
 \item Has your idea about radiation changed after finishing our
       project?
\end{enumerate}
  }
}
\\
\\

Answers by the students are 
summarized in table \ref{answers}. 
In most questions, answers from three students 
are essentially common and coincide with teacher's expectation.  
In Q.~1, all students agreed that the freedom allowed to them was
the exclusive feature of the project that is very different from
usual class.  In Q.~2, all students claimed the importance of planing 
research schedule in order to lead their collaboration
to success.  They also mentioned that their communication
skills were enhanced through interactions between project members and
opportunities of conference talks.  For example, they said that it 
was not easy to make appointment with each other.  They have learned
that it is important to manage their schedule. 
In Q.~4, Two students were satisfied with their success of developing their
cloud chamber.  Answers to Q.~5 are very impressive, because all students
agreed that communication with other members was major difficulty 
they confronted.  They implied  a small trouble between them during    
research, also they have not told its content.  
For Q.~6, they also told
that they need more knowledge and academic skill for realistic 
research.  They also agreed that the experience of an international
conference helps them to understand that English is 
necessary for scientific research in Q.~7.  
In Q.~8, they said that they
understood that the risk of radiation essentially depends on 
its amount.   All of them agreed that it is wrong to be afraid of 
radiation without  considering its amount and kind. 
\normalsize
\begin{table}[htbp]
\caption{Classification of answers to questions. In all questions, 
answers from three students can be classified to one or two kinds, which we 
denote answer 1 and 2. 
 }
\label{answers} 
 \begin{tabular}{cp{3cm}p{3cm}}
\hline
  Q.\ & Answer 1 & Answer 2    \\ \hline
  (1) & Making decision on their own  & \\ 
  (2) & Skill to plan research & Communication skill  \\ 
  (3) & Same as Q. (1)   & \\ 
  (4) & Success in their observation & \textit{Not so much} \\ 
  (5) & Relationship with other members & No time for meeting \\ 
  (6) & Amount of knowledge & Critical comments \\ 
  (7) & Important & Valuable experiences  \\
  (8) & Not so dangerous & \textit{Not changed so much}  \\ \hline
 \end{tabular}
\end{table}
There are two answers that contradict with  teacher's 
expectation.  They are shown in \textit{italic}  in Table
\ref{answers}.  Both are from a student who most contributed to and 
lead the entire project with very strong motivation.  For Q.~7, 
he answered that he surprised only little with 
the first success of observation, since it can be expected from
the theory and literature.  He also claimed that 
he had only gained a little knowledge of radiation 
since our project is rather focused on 
the development of a chamber than the physical aspects of radiation.  
These two issues were intended  by the teacher to lead our project
to success in limited time.  It is clear that he realized that teacher's 
intention, had more interests with physics itself, and
wanted to achieve a higher goal than the teacher expected.  
As is clear from this example, students has begun to think critically both for
their research plan and also physics itself.  

\subsection*{Discussions}

From the interviews with students, it turns out that providing an
opportunity of realistic research is very effective to promote 
their motivation for science and to develop technical and social skills
required for scientific research.  Although our program depends on
large support from SSH,  some  aspects will be useful for non-SSH
schools.   Opportunities for local conferences will available at
relatively low cost.   Finding enough time for research will be
very difficult except for science club students.  However, it is 
still possible to design inquiry based activity within usual physics
class, although time will be very limited.  Assigning student 
research project during summer vacation is also a good idea.  
In any case, giving enough freedom to students to choose their subject
 is important.

\section*{Acknowledgments}

We would like to thank M.~Sinta, S.~Hosoya and R.~Yamaishi for
support and encouragements and J.~Cooper for proofreading. 
We also like to thank
H.~Kanda, N.~Kanda, M.~Hayashi, K.~Taniguchi, J.~Yasuda and M.~Taniguchi 
for useful comments at JPS junior session on Mar. 2013.  
This work is supported by the Saito Kenzo honour fund and the SSH
project from the Ministry of Education, Culture, Sports, Science and Technology (MEXT) 
 and Japan Science and Technology agency (JST).

\end{document}